\documentclass[sigconf]{acmart}
\usepackage{graphicx}
\usepackage{epstopdf}
\usepackage{graphics}
\usepackage{url}
\usepackage{multirow,array}

\AtBeginDocument{%
  \providecommand\BibTeX{{%
    \normalfont B\kern-0.5em{\scshape i\kern-0.25em b}\kern-0.8em\TeX}}}

\acmConference[Arxiv]{Arxiv Preprint}
\begin{document}

\title{Fund2Vec: Mutual Funds Similarity using Graph Learning}

\author{Vipul Satone}
\email{vipulsatone@gmail.com}
\affiliation{%
  \institution{The Vanguard Group, Inc.}
}
\author{Dhruv Desai}
\email{dhruvdesai@alumni.upenn.edu}
\affiliation{%
  \institution{The Vanguard Group, Inc.}
}
\author{Dhagash Mehta}
\authornotemark[1]
\email{dhagashbmehta@gmail.com}
\affiliation{%
  \institution{The Vanguard Group, Inc.}
}

\renewcommand{\shortauthors}{Satone et al.}

\begin{abstract}
  Identifying similar mutual funds with respect to the underlying portfolios has found many applications in financial services ranging from fund recommender systems, competitors analysis, portfolio analytics, marketing and sales, etc. The traditional methods are either qualitative, and hence prone to biases and often not reproducible, or, are known not to capture all the nuances (non-linearities) among the portfolios from the raw data. We propose a radically new approach to identify similar funds based on the weighted bipartite network representation of funds and their underlying assets data using a sophisticated machine learning method called Node2Vec which learns an embedded low-dimensional representation of the network. We call the embedding \emph{Fund2Vec}. Ours is the first ever study of the weighted bipartite network representation of the funds-assets network in its original form that identifies structural similarity among portfolios as opposed to merely portfolio overlaps.
\end{abstract}

% <concept_id>10010405.10010455.10010460</concept_id>
% <concept_desc>Applied computing~Economics</concept_desc>

% \ccsdesc[500]{Applied computing~Economics}

%%
% Keywords. The author(s) should pick words that accurately describe
% the work being presented. Separate the keywords with commas.

\maketitle

\section{Introduction}

With the surge of popularity of the mutual funds and exchange-traded funds (ETFs), many investment managers have launched multiple funds with different investment strategies and philosophies yielding a plethora of funds available in the market. Such wide variety of available products in the market bring a classic business problem: identifying similar (and dissimilar) products. Identifying similar funds have multiple applications such as helping sales representative recommend \cite{aggarwal2016recommender} similar funds to the ones the investor has in their portfolio while bringing in additional advantages such as lower expense ratio, brand name, etc.; recommending complementing items to the ones the investor is buying based on what other investors bought together; in devising a tax loss harvesting strategy; analysing diversification within a portfolio of funds, portfolio analytics, etc.

Even with its wide applicability, rigorously quantifying 'fund similarity' is a highly complex problem as it may involve emotional aspects and personal biases, in addition to multiple technical challenges such as defining an objective metric of similarity, identifying the variables to suite the chosen definition of similarity as well as potential nonlinear relationships between the chosen variables.

In the financial domain, one of the most popular ways to find similar funds is to look up their respective fund categorization provided by third-party data vendors such as Morningstar \cite{morningstarcategorization} and Lipper \cite{lipperclassification}. The fund categorizations are recommended by certain committees of experts who rely on various quantitative and qualitative aspects of individual funds. However, with these categorizations, an investor can only get a list of similar funds to a chosen one, but not a ranking (i.e., which of the ones out of the list are more similar to the chosen one than the others in the same list). 

Various data-driven approaches for fund similarity have been proposed in the past \cite{marathe1999categorizing,sakakibara2015clustering, haslem2001morningstar}. Here, one uses raw data for each mutual fund in the selected universe of funds, and compute the Euclidean distance, cosine similarity scores or Jaccard index in the high-dimensional space of preselected variables. In more advanced investigations, they also used different unsupervised clustering techniques \cite{cai2016clustering} on the preselected variables to determine clusters of similar funds, and they compared these clusters with the respective Morningstar categorizations to identify mismatches \cite{orphanides1996compensation,brown1997mutual,dibartolomeo1997mutual,elton2003incentive,kim2000mutual,castellanos2005spanish,moreno2006self,acharya2007classifying,haslem2001morningstar,lamponi2015data} with the data-driven clustering though it turned out that the reason of mismatch was lack of important variables and improper use and interpretation of the data-driven approaches \cite{mehta2020machine,haslem2001morningstar} (and in a published comments by Gambera, Rekenthaler and Xia in \cite{haslem2001morningstar}).

Here, we aim to go beyond third-party categorizations in search for a data-driven approach to provide not only an objective way of identifying similar funds as possible, but also to come up with a similarity ranking system. To that end, we argue that manually choosing any aggregate level fund composition related variables (e.g., \% equity, \% fixed income, \% allocation in specific sectors) will invariably add certain inherent bias such as rigidly classifying certain companies in specific sector (e.g., Amazon.com Inc. being rigidly classified as strictly in retail, or technology sector). 

In the present work, we focus on the asset level information for each fund from a chosen universe: we view the fund and their underlying assets data as a weighted bipartite network. Then, we translate the fund similarity problem in the network science language. Finally, we tailor-make a sophisticated machine learning algorithm called Node2Vec to capture underlying nonlinear relationship within the data to identify similar funds.

\begin{comment}
The remainder of the paper is organized as follows: in Section \ref{sec:Previous_literature}, we review the existing literature on fund similarity and network science approaches for mutual funds and their underlying assets. In Section \ref{sec:fund_asset_network}, we provide a description of our data collection process and explain how the data is translated into network science terminology. In Section \ref{sec:method}, we describe the Node2Vec method using a simple example network. In Section \ref{sec:Results}, we device novel metrics to compare different representations of the data and show our results. Finally, in Section \ref{sec:conclusion}, we discuss the results and their implications, and conclude. 
\end{comment}

\section{Network of Funds and Assets}\label{sec:Previous_literature}

\begin{figure}[h]
    \centering
    \includegraphics[width=\linewidth]{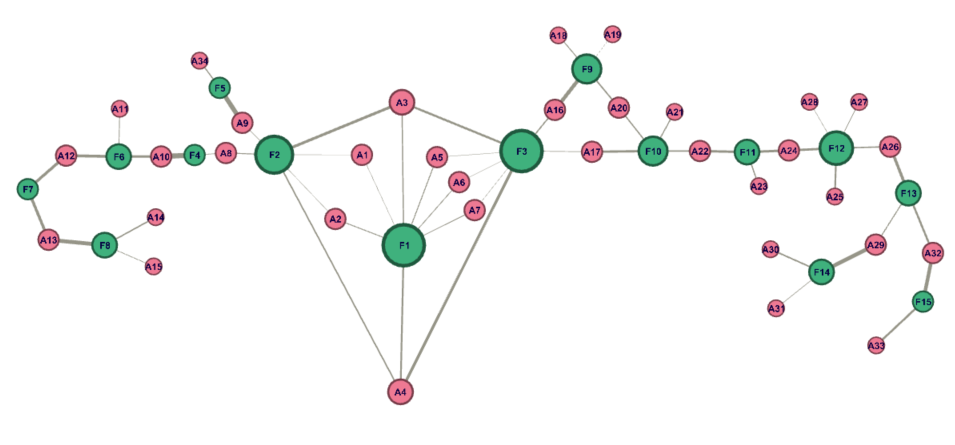}
    \caption{A hypothetical example of funds and assets network where green and red nodes correspond to funds and asset nodes, respectively. The thickness of links is proportional to the weight of the asset in the linked fund.}
    \label{fig:dummynetwork}
\end{figure}

We begin by viewing the mutual funds and their underlying holdings data as a weighted bipartite network where each mutual fund as well as each asset are represented as individual nodes. The link, or lack of it, between each fund-asset pair of nodes yields that the fund contains the asset at the specific snapshot in time. The weight on each link represents the percentage weight of the asset in the respective fund at the snapshot. Figure \ref{fig:dummynetwork} shows a simple example of fund-asset network with dummy data where $F1, \dots, F15$ are hypothetical funds and $A1, \dots, A33$ are their hypothetical assets.

Mutual funds and their underlying assets have been analysed from the network point of view in the past \cite{allen2009networks,d2016complex}. In \cite{solis2009visualizing}, a network of stocks and mutual funds for a selected list of total 18 stock-only mutual funds was investigated. For each of the 18 funds, the top 10 stocks according to the percentage weight in the fund were considered. This network of 18 funds and 99 stocks was shown to exhibit small-world characteristic. Here, the original bipartite network was projected into an unweighted unipartite network whose only nodes are stocks, and two stocks are connected if they belonged to the same mutual fund. The large average clustering coefficient (a quantity to analyze local structure of each node) \cite{watts1998collective} of the network was interpreted as higher probability than by chance for any two stocks to be in the same mutual fund if they both are present in another common mutual fund(s).

% In \cite{gualdi2016statistically}, networks between institutions (US based brokers, hedge funds, investment advisers, mutual funds, pension funds, private banks and others) and securities were studied using Factset Ownership database for duration between 1999Q1 to 2013Q4. There, the bipartite networks between institutions and securities were first binarized (i.e., all non-zero weights are assumed to have unit weight) and then projected onto the corresponding weighted unipartite networks using a tailor-made statistical significant weight. 

In \cite{mitali2019common} (and in \cite{lin2019identifying}), a similar projection was used on the weighted bipartite networks of US based mutual fund and assets with weights being total net assets, for duration from 1980Q1 to 2016Q4, first to binary bipartite network (i.e., all non-zero weights are assumed to have unit weight) and then projected to weighted unipartite networks where the only nodes are mutual funds. There, with the help of weighted degree centrality and weighted eigen centrality, the author concluded that the degree centrality has a negative and statistically significant effect on mutual fund performance, i.e., mutual funds with low portfolio similarity with their peers perform better. More recently, such networks corresponding to pension funds (Italian market) are also analyzed and found exhibiting similar characteristics as aforementioned networks \cite{d2019complex}.

In \cite{sakakibara2015clustering}, a network of 551 Japanese mutual funds was constructed using top 10 stocks of each fund. Here, the bipartite network was projected onto a unipartite network of funds where the weights between nodes were the number of common stocks. Then, the weighted network was clustered using the k-means and spectral clustering method for graph partitioning.

\subsection{Bipartite Networks vs Projected Networks}
Traditionally, bipartite networks have been investigated only after projecting it onto a unipartite network with one of the two types of nodes: the weighted bipartite networks exhibit certain peculiar characteristics compared to regular networks and hence many of the network quantities defined for unipartite networks may not be directly appropriate for weighted bipartite networks. Hence, most of the existing literature on mutual funds and assets networks have been limited to projected networks.  However, the original weighted bipartite network encodes the complete information about the underlying data whereas there is invariably some loss of information during any projection of bipartite network to a unipartite network \cite{borgatti1997network, borgatti20092,borgatti2011analyzing,latapy2008basic} though certain projections may retain more information (e.g., Ref.~\cite{zhou2007bipartite,lavin2019modeling}) than others. In general, such a projection typically induces a huge number of spurious edges which limits on the computations of different quantities in practice. Moreover, such large number of edges in the projected network may yield unique properties such as high clustering coefficient which may not be representative to corresponding characteristics of the original network. 

In \cite{delpini2019systemic}, the networks of mutual funds and assets for the US based mutual funds were investigated keeping the bipartite network structure intact for 2006Q3, 2007Q3, 2008Q3, i.e., before, during and after the 2008 financial crisis. The authors showed that the degree distribution of the fund-nodes as well as asset-nodes both exhibited scale-free characteristics. Here, the definition of degree was that of an unweighted bipartite network. They also showed that while individual mutual funds have become more diversified portfolios after the financial crisis, the mutual funds have become more similar to each other, i.e., there is a large overlap (calculated using asset weights within individual portfolios) among mutual funds, giving rise to systematic risk.

% Another work that comes closer to our approach at least in the sense of learning embedded representation of the mutual funds data (though it does not utilize the network representation of the data) is Ref.~\cite{moreno2006self} where a nonlinear clustering technique called self-organizing map (SOM) was used to learn lower dimensional representation of 1592 mutual funds from the Spanish market over ten fund attributes (average return, standard deviation, skewness, kurtosis, the 5\% of maximum losses, the 5\% of maximum returns, the reward-to-semivariability ratio, the beta against a chosen index, the beta against a 10-year notional bond and the correlation of each fund with an equally weighted benchmark obtained from each of the 14 legal categories in the Spanish market). Then the funds were clustered in the learned representation concluding that this clustering had significantly less missclassifications with respect to the corresponding legal categories compared to the one using the K-means clustering on the original raw variables.

In the present work, we analyze the fund-asset network in its original weighted bipartite network form without any approximation nor any projection. Then, the fund similarity problem translates into that of finding similar fund nodes on the network. We apply a recently proposed machine learning algorithm called Node2Vec \cite{10.1145/2939672.2939754} to obtain a lower dimensional representation of the high-dimensional network data, called embedded representation of the network. The lower dimensional representation captures most of the variance in the data, and more importantly the nonlinear relationships among the raw input features. We then perform similarity computation, such as cosine similarity, in the embedded representation to obtain similarity scores among different pairs of funds that are a result of nonlinear relationship of the features. 

\section{Fund-Asset Network and Data}\label{sec:fund_asset_network}
In order to have an entire universe of funds and assets associated with them, we scraped data from the US Security and Exchange Commission (SEC) filings for form NPORT-P retrieved from the SEC Electronic Data Gathering, Analysis, and Retrieval (EDGAR) database, which is the primary system for submissions by companies and others who are required by law to file information about their funds with the SEC. Access to EDGAR's public database is free. Under the SEC regulations funds must report their portfolio and each of their portfolio holdings as of last business day or last calendar day of the month. These reports disclose portfolio information as calculated by the fund for the reporting period’s ending net asset value, reported on Form N-PORT must be filed with the commission no later than 30 days after the end of each month. Information reported on Form N-PORT for the third month of each fund's fiscal quarter is made publicly available 60 days after the end of the funds fiscal quarter. Form N-PORT provides the following information regarding the funds: total assets, including assets attributable to miscellaneous securities reported for each underlying security, total liabilities, net assets, certain assets and liabilities reported in US dollars, securities lending, returns information, flow information, percentage value compared to net assets of the fund, etc.

Based on the filing data, all funds were filtered down to a timelinse falling in Quarter 1 of 2020. In order to have a defined universe we chose to subset the data to all Equity Index Funds. SEC provides a list of funds, their share class as Class IDs, and the Series ID which map the same fund of different share class to a single ID.

We create the bipartite network with each fund and assets as nodes and percentage investment of a fund in an asset as weight on the edge between the fund nodes. There are no edges between two funds or two assets as we only consider funds that do not invest in another funds, and assets do not invest in another asset. We also do not consider the direction of investment from fund to asset. Hence, we get an undirected weighted bipartite network. 

\subsection{Data Cleaning}
The data collected was cleaned to keep only edges with non-negative weights, and assets with a proper ISIN. Only those funds with at least 95\% of the portfolio is present were retained. Certain funds, mostly due to the missing data, were not connected to the giant connected fund-asset component and were removed. The data after this pre-processing contained about 1093 funds and 16,138 assets, and the basic network statistics is as shown in Table \ref{tab:network_basic_stat}.

\begin{table}[]
\begin{tabular}{|c|c|c|}
\hline
& \textbf{Fund Nodes} & \textbf{Asset nodes}     \\
\hline
\textbf{Count} & 1093  & 16,138  \\ \hline
\textbf{Mean number of edges} & {431.27447} & {29.20950}  \\ \hline
\textbf{Median number of edges} & {159} & {13} \\ \hline
\end{tabular}
\caption{Basic network statistics for the fund-asset network.}
\label{tab:network_basic_stat}
\vspace{-6mm}
\end{table}

\section{Methodology}\label{sec:method}
After its publication in 2016, Node2vec \cite{10.1145/2939672.2939754} has become one of the most used algorithms to learn lower dimensional representation for nodes in graph. Node2Vec is based on a word embedding technique called Word2Vec, hence, we first briefly describe Word2Vec.

\subsection{Word2Vec}
Word2vec \cite{NIPS2013_5021} is one of the most widely used word embedding techniques in the natural language processing (NLP) tasks. Instead of the traditional methods to encode words in a large corpus of text in the numerical vectors form such as one-hot-encoding, where the vector length would be equal to vocabulary size and each element of the vector represents a word in the vocabulary, Word2Vec constructs a lower dimensional space that captures meaningful semantic and syntactic relationships between words.  

Here, first, each sentence is viewed as an directed subgraph where each word is a node of the graph corresponding to the sentence. Then, a shallow two-layer neural network is then used to get Word2Vec embeddings. Input to Word2Vec is large corpus of sentences, i.e., sequence of words, and it outputs a vector space where each word is represented by a unique vector. In the vector space, the words which share a common context in the corpus lie close to each other. Word2Vec uses one of the two architectures, namely 'Continuous Bag-of-Words (CBOW)' or 'Skip-gram' model to find word embedding. 
 
\noindent\textit{CBOW:} In CBOW model, the surrounding context words are used to predict the target word. 
 
\noindent\textit{Skip-Gram:} Here, the target word is fed as input, whereas the context is generated as the output.

\subsection{Node2Vec}
Though Word2Vec very efficiently embeds data consisting of directed subgraphs, in other applications than NLP, graphs may arise in different flavors such as (un)directed, (un)weighted, (a)cyclic, and hence the methodology of Word2Vec cannot be directly applicable to embed these graphs.

Node2vec \cite{10.1145/2939672.2939754} solves this problem by using a clever trick that 'generates' directed subgraphs from the other types of graphs by starting from each node of the original graph and generating random walks. The set of nodes in the original graph is then considered as the 'vocabulary' and the directed path of each random walk is considered as a 'sentence'. This new data, that now resembles the textual data, can be fed into Word2Vec to finally obtain the desired embedding. This algorithm returns a feature representation that maximizes the likelihood of preserving network neighbourhoods of nodes in a $d$-dimensional feature space. 

The output embeddings or representations will depend on network neighbourhoods, and the sampling of these network neighbourhoods (i.e., the random walks) is a very important task. Below we list the most important hyperparameters for Node2Vec:
\begin{enumerate}
\item The Number of random walks ($r$): the number of random walks to be generated from each node in the graph;
\item The length for each random walk ($l$): the length (number of hops) of each random walk from each node in the graph;
\item $p$: the probability with which a random walk will return to the node it already visited previously; and,
\item $q$: the probability with which a random walk will explore the unexplored part of the graph.
\end{enumerate}
Note that these hyperparameters are specific to Node2Vec, in addition to the hyperaparameters such as context window size, number of iterations for the Word2Vec algorithm.

Node2vec can be fine tuned to conform to already established equivalence in network science and can interpolate between the breadth-first sampling (BFS) (where nodes are sampled from the immediate neighbourhood of the starting node) and depth-first sampling (DFS) (where nodes are sequentially sampled from increasing distance from the source), i.e., between homophily (nodes that are highly connected should have their embedding vectors close to each other) and structural equivalence (nodes that have similar structural roles should be embedded closely together). 

\section{Experiments and Results}\label{sec:Results}
We aim to find the lower ($d$) dimensional embedding of the network data using Node2Vec such that the data-points corresponding to similar funds will be closer to each other. A systematic and objective evaluation of embeddings for an unsupervised technique is yet to be rigorously defined in the literature \cite{mara2019evalne}. Here, we device novel metrics to evaluate the embeddings for the application at hand and tune hyperparmaeters with respect to the new metrics. We call the thus obtained best embedding \textit{Fund2Vec}. 

\subsection{`Bipartiteness' as a Metric}

We begin with the observation that the network at hand is inherently bipartite, and any network learning algorithm should at least retain the `bipartiteness' of the network in the lower-dimensional representation. The embedded representation, however, is an abstract manifold and the data-points on this representation do not necessarily have network interpretation making it difficult to come up with a metric to measure the retained bipartiteness. 

Here, we propose a pragmatic approach: in our bipartite network, the only true ``labels`` in the data are ``fund`` and ``asset``. In the embedded representation should at least bring all the funds (assets) close to other funds (assets). We do not expect that in the embedded representation there will only be exactly 2 clusters, one consisting of funds and the other consisting of assets. Then, after training Node2Vec at each hyperparameter point, we employed the K-means \cite{macqueen1967} algorithm for a range of values for $K$ to cluster the data and then look for the clustering which most clearly distinguish funds from clusters, i.e., funds and assets are respectively clustered together. A goal of hyperparameter optimization is then to seek for a hyperparameter point at which the mismatch between clustering and the ground truth labels is minimized. In summary, we propose the bipartiteness as a metric to evaluate the embeddings learned by Node2Vec, or any other related graph learning method, when the original network is a bipartite network.

In practice, to measure how much bipartiteness is preserved by the given embedded representation, we measure the mismatch between the clustering from K-means and the ground truth labels (i.e., fund and asset) using a classic metric called the V-measure which takes into account both homogeneity and completeness of the cluster. Below, we recall the definitions of homogeneity, completeness and V-measure \cite{scikit-learn, rosenberg2007v}.

\noindent \textit{Homogeneity:} When each of the clusters only contains data points which are members of a single ground truth class, the clustering is called homogeneous. If a cluster has data-points from two or more different classes it is called a heterogeneous cluster.

\noindent\textit{Completeness:} Completeness is defined as the measure of a cluster when a cluster contains all the elements of a class, as opposed to a cluster that fails to capture one or more elements of the class.

\noindent\textit{V-Measure:} V-measure is then defined as the harmonic mean between the homogeneity and completeness, i.e.,\\
\hfill\\
\centerline{$ V-measure = \frac{  (1+\beta) (Homogeneity) (Completeness) }{ (\beta * Homogeneity + Completeness)}  $, }\\

where $\beta$ can be used to assign more weight to either homogeneity ( $\beta < 1$) or completeness ($\beta > 1$). Since the number of ground truth labels is only 2, clearly, the completeness will decrease as $K$ increases. We rather focus on maximizing homogeneity. To systematically weigh homogeneity significantly more than completeness in the computation of V-measure, we use $\beta = 0.01$. 
% \textcolor{red}{ Is any explanation needed for : Beta is very low and homogenity is highly favored. Why not we just use homogenity?}

% In addition to V-measure, we also use true positive (TP), false positive (FP), true negative (TN), false negative (FN), and accuracy as further metrics to evaluate the embeddings: here, the ground truth labels are Fund and Asset. We do not use true labels to determine fund or asset cluster. For each cluster, we take the majority vote for the ground truth labels from the data-points within the cluster, and then assign this label to the cluster (e.g., in a cluster of 1000 data-points, if 790 data-points are funds, then the cluster label is 'fund'). Then, TP, FP, TN, FN and accuracy are defined to compare the ground truth labels with the cluster labels.

\subsubsection{Hyperparameter Optimization}
With the above metrics, we perform hyperparameter optimization for Node2Vec to obtain the desired embedding. The hyperparameters we tune are $d$, $l$, $r$, $p$ and $q$. These hyperparameters explore various strategies to sample the network starting from interpolating between BFS and DFS, to smaller and larger dimensional embeddings. We did not extensively tune hyper-parameters of the underlying skip-gram model as our main aim in this investigation was to study the effects of different random sampling strategies on embeddings.

In summary, below is the hyperparameter optimization process we followed: For each hyperparameter point, we employ the K-means algorithmn (with Euclidean distance) for $K$ ranging from $2$ to $10$ (higher values of $K$ did not add further information in this case); we then compute the V-measure ($\beta = 0.01$) for each clustering with respect to the ground truth labels (i.e., fund and asset); then, we chose the hyperparameter point at which the V-measure attained the maximum value.

\begin{figure}[h]
    \centering
    \includegraphics[width=0.45\textwidth]{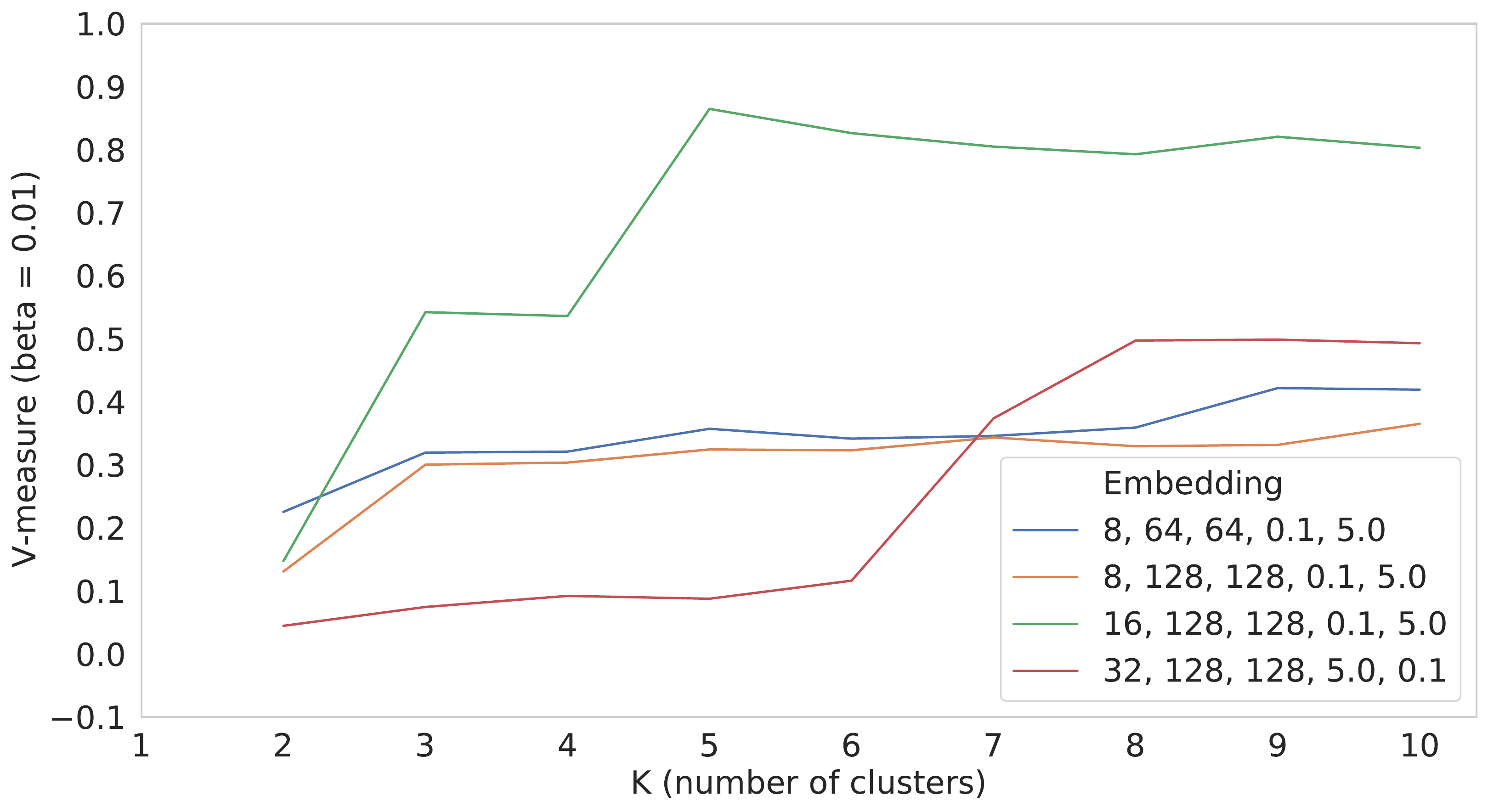}
    \caption{Plot for $K$ vs V-measure ($\beta=0.01$) for a few hyperparameter points.}
    \label{fig:hyperparameter_opt}
\end{figure}

Plots for a few values of hyperparameter points are shown in Figure (\ref{fig:hyperparameter_opt}). The results for optimal values of $K$ for each hyperparameter point in our experiments and the corresponding values of various metrics are shown in table \ref{table:tablethree}. For the data at hand, the best hyperparameter point in our experiments was: $d=16$, $l=128$, $r= 128$, $p = 0.1$ and $q = 5$. The V-measure for these parameters is $0.86$ for $K=5$, yielding a clear separation between funds and assets in the optimal embedding. From here on, we call this $16$-dimensional representation as the \emph{Fund2Vec}.Out of the five clusters, two of them only consists of funds the remaining cluster consists of a mixture of funds and assets with majority of data-points (>99 \%) corresponding to assets.

\begin{table}[]
\resizebox{\columnwidth}{!}{

\begin{tabular}{|l|l|l|l|l|l|l|}
\hline
d           & \textbf{l}   & \textbf{r}   & \textbf{p}   & \textbf{q} & \textbf{Optimal $K$} & \textbf{V-measure ($\beta   = 0.01$)} \\ \hline
8  & 128 & 128 & 0.1 & 5   & 10 & 0.365 \\ \hline
8  & 128 & 128 & 5   & 0.1 & 7  & 0.807 \\ \hline
8  & 64  & 64  & 0.1 & 5   & 9  & 0.421 \\ \hline
8  & 64  & 64  & 5   & 0.1 & 11 & 0.782 \\ \hline
\textbf{16} & \textbf{128} & \textbf{128} & \textbf{0.1} & \textbf{5} & \textbf{5}         & \textbf{0.864}                    \\ \hline
16 & 128 & 128 & 5   & 0.1 & 10 & 0.607 \\ \hline
16 & 64  & 64  & 0.1 & 5   & 10 & 0.829 \\ \hline
16 & 64  & 64  & 5   & 0.1 & 11 & 0.722 \\ \hline
32 & 128 & 128 & 0.1 & 5   & 10 & 0.801 \\ \hline
32 & 128 & 128 & 5   & 0.1 & 9  & 0.498 \\ \hline
32 & 64  & 64  & 0.1 & 5   & 9  & 0.782 \\ \hline
32 & 64  & 64  & 5   & 0.1 & 9  & 0.64 \\ \hline
\end{tabular}
}
\caption{The table shows the effect of the hyperparameters on various metrics. The best hyperparameter point from our runs is shown in bold text.}
\label{table:tablethree}
\vspace{-6mm}
\end{table}

\begin{figure}[h]
    \centering
    \includegraphics[width=0.45\textwidth]{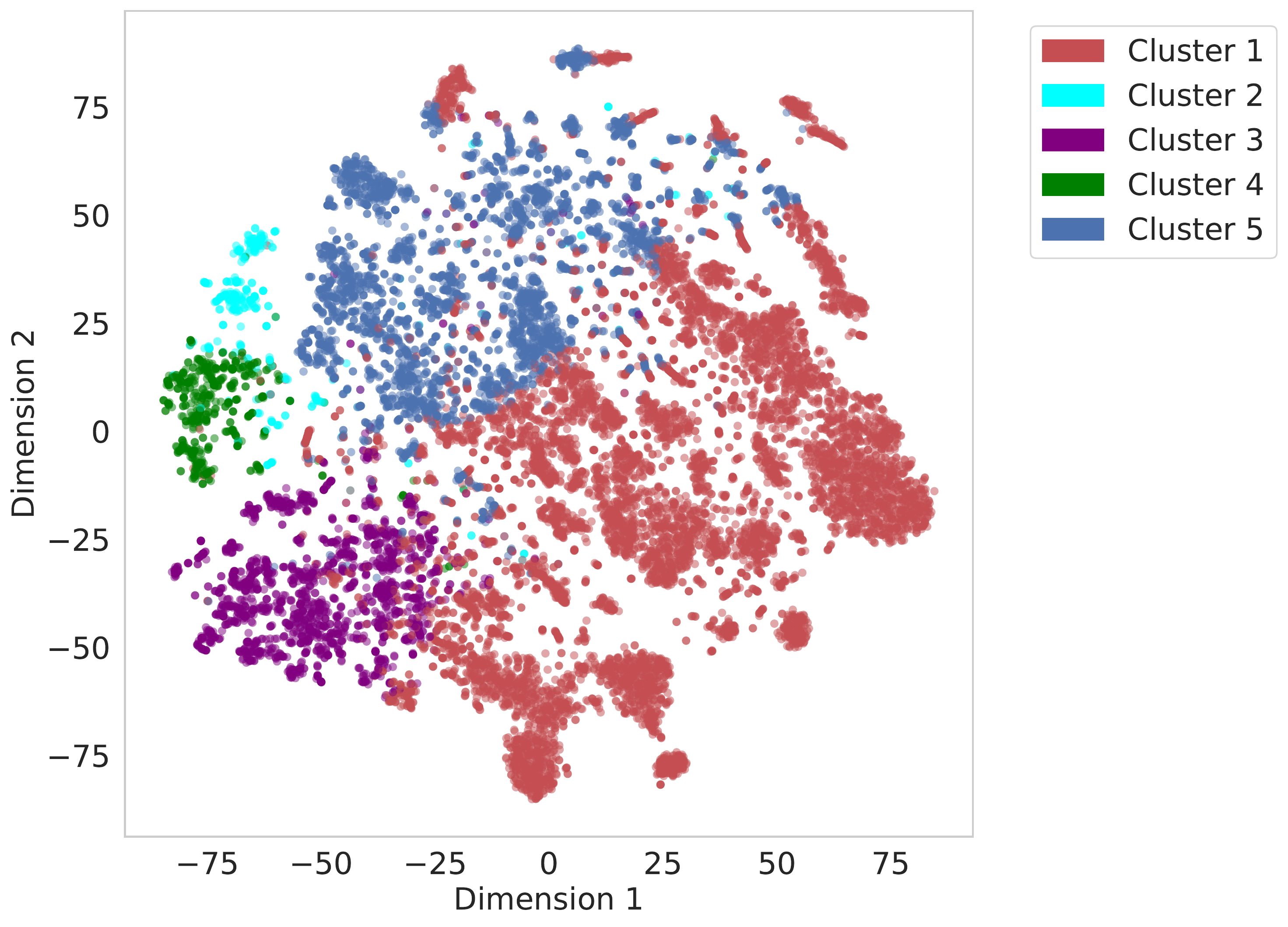}
    \caption{A TSNE plot of data-points in the optimal embedding. The blue cluster contains data points that correspond to asset nodes in the original network, whereas the red cluster contains data points that correspond to fund nodes.}
    \label{fig:clustering}
\end{figure}

Figure \ref{fig:clustering} shows the T-SNE plot for all the data-points in this optimal embedding. Where most of the fund and asset nodes are clearly in separate clusters. Cluster 2 and cluster 4 consist only of funds, in the T-SNE plot they lie very close to each other. The data-points which were \textit{missclassified} indeed correspond to special types of funds or assets as discussed in the next subsection.

%Following table \ref{table:tableone} is a sample result for fund fund\_668 : One important thing to notice here is that while we compute similarity using node2vec, we check for similar nodes from all the available 17,300 nodes (i.e fund nodes and asset nodes). Node2vec has calculated embeddings which differentiate between assets and funds so we have got all fund nodes similar to our query fund node fund\_668. However, when we calculate similarity using 16175 long vectors, we check for similar funds from 1125 fund nodes and not all 17300 nodes.

\subsubsection{Computational Details}
The computation took, on an average, around 6 hours to generate embeddings using Node2Vec on a regular single core machine with 60 GB RAM. We used Scikit-Learn \cite{scikit-learn} for the data pre-processing and K-means computation as well as to compute various evaluation metrics.  

\subsubsection{A few words on Node2Vec for bipartite networks}
In \cite{gao2018bine}, it was argued that Node2Vec (and other network embedding methods) focused on homogeneous networks and overlooks special properties of bipartite networks. The authors proposed a novel method which first generates node sequences that preserve the long-tail distribution of nodes in the original bipartite network, and then utilizes a novel optimization framework that accounts for both the explicit and implicit
relations in learning the embedded representations.

In the present work, while using Node2Vec we have devised the above bipartiteness metric that explicitly help tuning the model towards an embedding that preserves the properties of the original bipartite network. A comparison between our methodology, which is more ad-hoc and pragmatic, and the one proposed in \cite{gao2018bine} which extends the Node2Vec methodology from more theoretical point of view, is beyond the scope of this paper. 

\subsubsection{Analysis of miss-classified funds}
Table \ref{table:tablemissclassification} shows the number of funds and assets in each of the 5 clusters after K-means ($K=5$) clustering within \emph{Fund2Vec}. In total there are 67 funds which were ``misclassified`` as assets, i.e., these funds appeared in a cluster that mostly consists of assets otherwise. Interestingly, all these misclassified funds exhibited specific and rather unique characteristic with respect to the other funds: all these funds were either currency-hedged equity index funds which have assets from countries other than the US. The assets within these funds only rarely appeared in other funds than the currency-hedged funds making the currency-hedged funds create almost star-graph structure around them. In a future work, we will discuss this interesting set of \textit{outlier} funds.

\begin{table}[]
\resizebox{\columnwidth}{!}{
\begin{tabular}{|l|l|l|l|}
\hline
\textbf{Cluster number \#} & \textbf{\# Funds} & \textbf{\# Assets} & \textbf{Cluster Type}  \\ \hline
1              & 60 (0.60\%)   & 9933 (99.40\%) & Asset Cluster \\ \hline
2              & 383 (100\%)   & 0 (0.0\%)      & Fund Cluster  \\ \hline
3              & 1 (0.05\%)    & 2041 (99.95\%) & Asset Cluster \\ \hline
4              & 643 (100\%)   & 0 (0.0\%)      & Fund Cluster  \\ \hline
5              & 6 (0.14\%)    & 4164 (99.86\%) & Asset Cluster \\ \hline
\textbf{Total}          & 1093          & 16138          &               \\ \hline
\end{tabular}}
\caption{The table summarizes the distribution of funds and assets among all the 5 clusters obtained using $K$-means within \emph{Fund2Vec}.}
\label{table:tablemissclassification}
\vspace{-6mm}
\end{table}

\subsection{Baseline Benchmark}
Here, we devise another metric to perform sanity check to ensure the final embedding indeed respects another objective ground truth information: since the funds' filings also provide information about the benchmarks they track, \emph{Fund2Vec} is expected to place funds which track the same benchmark closer than other groups of funds.

We chose two of the most popular benchmarks, namely, S\& P 500 (27 funds) and Russell 2000 (16 funds) and computed (1) the mean cosine similarity, and (2) standard deviation of the cosine similarities among all the pairs of funds tracking the same benchmark, using the original representation as well as \emph{Fund2Vec}. We also computed both these quantities for cosine similarities between each fund tracking the benchmark and all other funds not tracking the benchmark. The results are shown in Table \ref{tab:benchmark} clearly yielding that, on an average, \emph{Fund2Vec} brought funds tracking the respective index closer together compared to funds outside the benchmarks. Note that the relatively lower mean cosine similarity score among funds within the Russell 2000 benchmark in the original space may be due to the curse of dimensionality, whereas \emph{Fund2Vec} being a low-dimensional representation evades this problem.

\begin{table*}[t]
\begin{tabular}{|r|l|l|l|l|l|l|l|l|l|}
\hline
\multirow{0}{*}{\textbf{Benchmark}} & \multirow{0}{*}{\textbf{\# Funds}} & \multicolumn{4}{l|}{\textbf{Fund2Vec}} & \multicolumn{4}{l|}{\textbf{Original Representation}} \\ 
\cline{3-10} & & \multicolumn{2}{l|}{\textbf{Within Benchmark}} & \multicolumn{2}{l|}{\textbf{Outside Benchmark}} & \multicolumn{2}{l|}{\textbf{Within Benchmark}} & \multicolumn{2}{l|}{\textbf{Outside Benchmark}} \\ 
\cline{3-10} &  & \textbf{Mean} & \textbf{Std Dev} & \textbf{Mean} & \textbf{Std Dev} & \textbf{Mean} & \textbf{Std Dev} & \textbf{Mean} & Std Dev \\ \hline
\multicolumn{1}{|l|}{S\&P 500}  & 27 & 0.99117 & 0.01307 & 0.29267 & 0.33861 & 0.96891 & 0.04357 & 0.17072 & 0.27734 \\ \hline
\multicolumn{1}{|l|}{Russell 2000}  & 9 & 0.94660 & 0.04633 & 0.12425  & 0.27225 & 0.38490 & 0.23163 & 0.03346 & 0.09417 \\ \hline
\end{tabular}
\caption{The table shows the mean and standard deviation of cosine similarities between funds tracking the same benchmark and those with funds not tracking the same benchmark.}
\label{tab:benchmark}
\vspace{-6mm}
\end{table*}

\subsection{Similarities in Different Representations}
Though the above defined bipartiteness metric provides a quantitative measure to evaluate embeddings with respect to the underlying network structure, it does not necessarily yield a 'goodness' of the embeddings in terms of similarity of funds. For an unsupervised problem such as the present one, one may not even expect a unique and objective definition of similarity to begin with. Here, we focus on a few reasonable and mathematically rigorous definitions that suits our purpose: cosine similarity and Jaccard index. 
Then, following the approach proposed in \cite{liu2019hood2vec}, instead of evaluating the goodness of the new embedding compared to the original representation of the data, we tailor-make a few metrics to ensure that the new embedding has indeed learned a \textit{different} view of the data and, in turn, of similarity.

\subsubsection{Cosine Similarity}
The cosine similarity is one of the most popular scores to measure similarity for multi-dimensional data. Here, each data point is considered as a point in the $n$-dimensional space where $n$ is the number of variables or features. Then, the cosine similarity between the $i$-th and $j$-th funds as 
\begin{equation}
    C_{i,j} = \frac{\textbf{W}_{F_i} . \textbf{W}_{F_j}}{\left\lVert\textbf{W}_{F_i} \right\rVert  \left\lVert\textbf{W}_{F_j} \right\rVert},
    \label{eq:cosine_similiarity}
\end{equation}
where $\textbf{W}_{F_i}$ and $\textbf{W}_{F_j}$ are vectors whose elements are weights of all the assets within funds $F_i$ and $F_j$, and $\left\lVert . \right\rVert $ is the Euclidean norm. $C_{i,j}$ ranges from $-1$ (i.e., $F_{i}$ and $F_{j}$ are completely dissimilar) to $1$ (i.e., $F_{i}$ and $F_{j}$ are completely identical).

Here, we compute the cosine similarity for the following two representations: 
\begin{enumerate}
    \item We compute the cosine similarity between each pair of funds in the original $16138$-dimensional asset-weights space. Hence, the corresponding $\textbf{W}_{F_i}$s in Eq.~(\ref{eq:cosine_similiarity}) are $16138$-dimensional vectors.
    \item We also compute the cosine similarity between each pair of funds in \textit{Fund2Vec}. Hence, the corresponding $\textbf{W}_{F_i}$s in Eq.~(\ref{eq:cosine_similiarity}) are $16$-dimensional vectors.
\end{enumerate}

\subsubsection{Jaccard Index}
Jaccard index is used to find similarity between two sample sets. It is defined as size (i.e., the number of elements) of the intersection between the two sets divided by the size of union of two sets. Jaccard index close to being $1$ indicates high similarity, and close to $0$ indicates little similarity between the two sets. 

Jaccard index between a pair of sets $A$ and $B$ is defined as 
\begin{equation}
    J_{A,B} = \frac{|A \cap B|}{|A \cup B|}.
\end{equation}
In other words, Jaccard index measures overlap between a pair of sets.
\subsubsection{A Metric to Compare Different Embeddings}
We compare the Jaccard indices using the above two scenarios for which the cosine similarity scores for each pair of funds are available. For each fund $F_i$, we query top $m = 5, 10, 20, 50$, most similar funds (with respect to the cosine similarity computation) in the $16138$-dimensional representation as well as in \textit{Fund2Vec}. Hence, for each $F_i$ we get two sets of similar funds corresponding to two representations, each set consisting of $m$ funds. Then, we measure the overlap between these two sets using Jaccard index.

For a fund, thus defined Jaccard index being close to $1$ for a value of $m$ means that the embedded representation is providing the same list of similar funds as the original representation, whereas the value close to $0$ means that the two representations are providing different views of similarity of the underlying data. Clearly, the larger the value of $m$ is, the greater are the chances of the same funds appearing in both the lists for a given fund with the extreme case being $m = n-1$, where $n$ is the total number of funds.

\begin{figure*}[h]
    \centering
    \includegraphics[width=\linewidth]{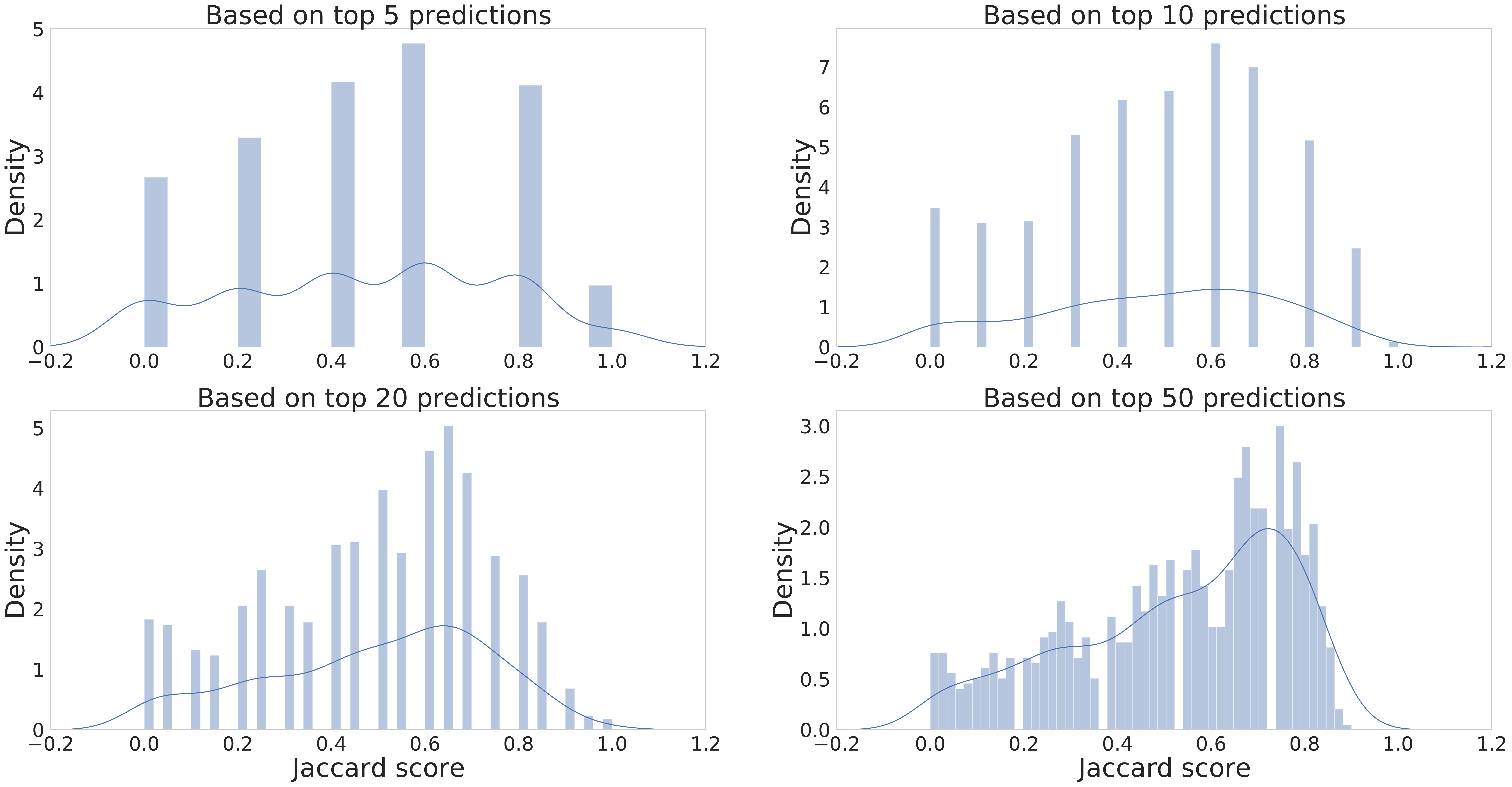}
    \caption{Statistics of Jaccard index for different values of $m = 5, 10, 20, 50$.}
    \label{fig:euclidean_jaccard_compare}
\end{figure*}

The distributions of Jaccard index over all the funds for each value of $m$ is shown in the figure \ref{fig:euclidean_jaccard_compare}, and further statistics for each distribution is summarized in Table \ref{tab:jaccard_stat}. The mean of Jaccard index for different values of $m$ ranges from 0.47 to 0.53 yielding, on an average, a relatively low overlap between list of similar funds for each fund. Figure \ref{fig:euclidean_jaccard_compare} demonstrates low overlaps in lists of similar funds between the two representations for smaller values of $m$.

\begin{table}[h]
\resizebox{\columnwidth}{!}{\begin{tabular}{|l|l|l|l|}
\hline
\textbf{Number of  funds querried (m)} & \textbf{Mean} & \textbf{Median} & \textbf{Standard   deviation}    \\ \hline
\textbf{5}  & 0.47 & 0.4  & 0.29 \\ \hline
\textbf{10} & 0.48 & 0.5  & 0.25 \\ \hline
\textbf{20} & 0.49  & 0.55 & 0.24 \\ \hline
\textbf{50} & 0.53 & 0.58 & 0.23 \\ \hline
\end{tabular}}
\caption{Table for Jaccard index that measures the overlap, for each fund $F_i$, between the top $m$ most similar funds to $F_i$ in the original representation and \textit{Fund2Vec}.}
\label{tab:jaccard_stat}
\vspace{-6mm}
\end{table}

\subsubsection{Cosine Similarities and Pearson Correlation}

\begin{figure}[h]
    \centering
    \includegraphics[width=0.5\textwidth]{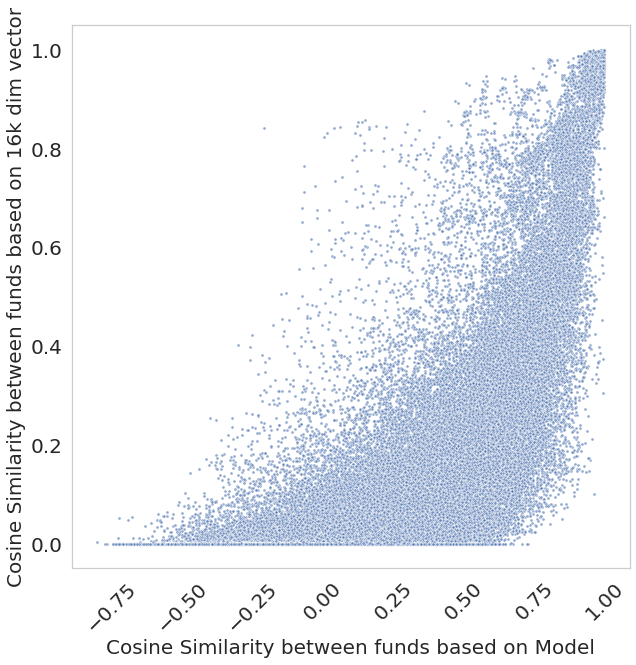}
    \caption{Pearson correlation comparison}
    \label{fig:pearson}
\end{figure}

Another way to find if we are able to capture a different view on similarity is to use the Pearson correlation between the two sets of similarity values. For every pair of funds, we calculated cosine similarity in the 16138 dimensional representation as well as within the Node2Vec embedding. If Node2Vec did not capture any new information than the original representation, then these cosine similarity values should be highly correlated. In our case, the average Pearson correlation is 0.66, which indicating that \emph{Fund2Vec} indeed attempted to capture potentially a different version of similarity (apparently, the structural similarity) among funds. Figure \ref{fig:pearson} shows the scattered plot of the cosine similarities for each pair of funds based on both the methods and exhibits a non-linear relationship between both the cosine similarities.

\section{Discussion and Conclusion}\label{sec:conclusion}
Due to the recent popularity of mutual funds and ETFs, a wide variety of mutual funds have been available in the market. In this paper, we pose the problem of comparing different mutual funds as a similarity problem on a corresponding weighted bipartite network where the network consists of funds and their underlying assets as nodes and weights of individual asset in given fund as the weights on the links between the fund and the asset. Ours is the first study providing an investigation of the funds-assets network while retaining the weighted bipartite characteristic of the network completely intact.

Traditionally similarity of nodes of a network is defined with respect to various quantities such as node centrality, degree centrality, eigenvector centrality, clustering coefficients etc. Instead of hand-crafting and computing such (potentially, infinitely many) quantities, we employ a machine learning methodology which yields a lower-dimensional manifold which captures the nonlinear relationship among nodes and links. Then, the cosine similarity between each pair of funds in the lower-dimensional representation provides an objective and data-driven definition of fund similarity. In particular, we employ a recently proposed network embedding technique called Node2Vec, prompting the name of embedding of funds-asset networks, \emph{Fund2Vec}.

Evaluating performance of embeddings in representation learning paradigm is an open problem. In this paper, we also develop domain knowledge based methodology to evaluate the quality of embeddings. Specifically, as a start, since the underlying network is a bipartite network, we used bipartiteness as our first metric to evaluate the quality of embedding: if the K-means clustering, for $K$ raning from 2 to 10, performed within the given embedding is able to cluster fund nodes and asset nodes into separate clusters, then the embedding preserves the original bipartiteness of the original network. The embeddings that mix fund nodes with asset nodes, and vice versa, when the K-means clustering is performed, are worse than the former embedding. We used this metric for hyperparameter optimization.

As a sanity check with respect to available objective ground truth information, we also computed cosine similarities between funds tracking the same index (S\& P 500 and Russell 2000) to demonstrate that \emph{Fund2Vec} indeed place these funds closer to each other compared to other group of funds.

We also used Jaccard metric and Pearson correlation to demonstrate that \emph{Fund2Vec} indeed provides a \textit{different} representation of data and, more importantly, different similarity scores. In particular, \emph{Fund2Vec} not only (implicitly) captures the similarity measures such as the portfolio overlaps between the pairs of funds, but, more importantly, it also captures structural similarities between funds, i.e., two mutual funds are similar if they have similar network structure. To the best of our knowledge, this is the first work that considers structural similarity of mutual funds and assets networks, while also saving time to manually creating individual network quantities to define structural similarity by hand.

Another benefit that \emph{Fund2Vec} provides over the other methods is that it considers the entire network while finding out embedding for each node. As a result we find similar funds even if they have less nodes in common. We anticipate that \emph{Fund2Vec} may have many more applications in addition to the ones mentioned in the paper due to the captured structural similarity, e.g., \emph{Fund2Vec} may also be used to help a portfolio constructor identify funds similar to the given one but only from within a certain theme such as Environment, Social and Governance (ESG), retail, fintech, etc.
\section*{Acknowledgement} The work presented here is a result of a pure and exploratory research work by the authors, and the authors are solely responsible for any mistakes and not The Vanguard Group. The authors would like to thank Victor Allen, James Belasco, Eduardo Fontes, Richa Sachdev and Hussain Zaidi for their feedback.

\hspace{0.6cm}

\textbf{Notes:} All investing is subject to risk, including the possible loss of the money you invest. Diversification does not ensure a profit or protect against a loss.

\copyright  2021 The Vanguard Group, Inc. All rights reserved.
\bibliography{main}{}

\begin{thebibliography}{10}

\bibitem{aggarwal2016recommender}
Charu~C Aggarwal et~al.
\newblock {\em Recommender systems}.
\newblock Springer, 2016.

\bibitem{morningstarcategorization}
Morningstar categorization.

\bibitem{lipperclassification}
Lipper u.s. fund classification, 2020.

\bibitem{marathe1999categorizing}
Achla Marathe and Hany~A Shawky.
\newblock Categorizing mutual funds using clusters.
\newblock {\em Advances in Quantitative analysis of Finance and Accounting},
  7(1):199--204, 1999.

\bibitem{sakakibara2015clustering}
Takumasa Sakakibara, Tohgoroh Matsui, Atsuko Mutoh, and Nobuhiro Inuzuka.
\newblock Clustering mutual funds based on investment similarity.
\newblock {\em Procedia Computer Science}, 60:881--890, 2015.

\bibitem{haslem2001morningstar}
John~A Haslem and Carl~A Scheraga.
\newblock Morningstar's classification of large-cap mutual funds.
\newblock {\em The Journal of Investing}, 10(1):79--89, 2001.

\bibitem{cai2016clustering}
Fan Cai, Nhien-An Le-Khac, and Tahar Kechadi.
\newblock Clustering approaches for financial data analysis: a survey.
\newblock {\em arXiv preprint arXiv:1609.08520}, 2016.

\bibitem{orphanides1996compensation}
Athanasios Orphanides et~al.
\newblock {\em Compensation incentives and risk taking behavior: evidence from
  mutual funds}.
\newblock Citeseer, 1996.

\bibitem{brown1997mutual}
Stephen~J Brown and William~N Goetzmann.
\newblock Mutual fund styles.
\newblock {\em Journal of financial Economics}, 43(3):373--399, 1997.

\bibitem{dibartolomeo1997mutual}
Dan DiBartolomeo and Erik Witkowski.
\newblock Mutual fund misclassification: Evidence based on style analysis.
\newblock {\em Financial Analysts Journal}, 53(5):32--43, 1997.

\bibitem{elton2003incentive}
Edwin~J Elton, Martin~J Gruber, and Christopher~R Blake.
\newblock Incentive fees and mutual funds.
\newblock {\em The Journal of Finance}, 58(2):779--804, 2003.

\bibitem{kim2000mutual}
Moon Kim, Ravi Shukla, and Michael Tomas.
\newblock Mutual fund objective misclassification.
\newblock {\em Journal of Economics and Business}, 52(4):309--323, 2000.

\bibitem{castellanos2005spanish}
Arturo~Rodr{\'\i}guez Castellanos and Bel{\'e}n~Vallejo Alonso.
\newblock Spanish mutual fund misclassification: Empirical evidence.
\newblock {\em The Journal of Investing}, 14(1):41--51, 2005.

\bibitem{moreno2006self}
David Moreno, Paulina Marco, and Ignacio Olmeda.
\newblock Self-organizing maps could improve the classification of spanish
  mutual funds.
\newblock {\em European Journal of Operational Research}, 174(2):1039--1054,
  2006.

\bibitem{acharya2007classifying}
Debashis Acharya and Gajendra Sidana.
\newblock Classifying mutual funds in india: Some results from clustering.
\newblock {\em Indian Journal of Economics and Business}, 6(1):71--79, 2007.

\bibitem{lamponi2015data}
Daniele Lamponi.
\newblock A data-driven categorization of investable assets.
\newblock {\em The Journal of Investing}, 24(4):73--80, 2015.

\bibitem{mehta2020machine}
Dhagash Mehta, Dhruv Desai, and Jithin Pradeep.
\newblock Machine learning fund categorizations.
\newblock {\em ACM International Conference on AI in Finance 2020.}, 2020.

\bibitem{allen2009networks}
Franklin Allen and Ana Babus.
\newblock Networks in finance.
\newblock {\em The network challenge: strategy, profit, and risk in an
  interlinked world}, 367, 2009.

\bibitem{d2016complex}
Anna~Maria D’Arcangelis and Giulia Rotundo.
\newblock Complex networks in finance.
\newblock In {\em Complex networks and dynamics}, pages 209--235. Springer,
  2016.

\bibitem{solis2009visualizing}
Rafael Solis.
\newblock Visualizing stock-mutual fund relationships through social network
  analysis.
\newblock {\em Global Journal of Finance and Banking Issues}, 3(3), 2009.

\bibitem{watts1998collective}
Duncan~J Watts and Steven~H Strogatz.
\newblock Collective dynamics of ‘small-world’networks.
\newblock {\em nature}, 393(6684):440--442, 1998.

\bibitem{mitali2019common}
Shema~F Mitali.
\newblock Common holdings and mutual fund performance.
\newblock {\em Available at SSRN 3448494}, 2019.

\bibitem{lin2019identifying}
Li~Lin and Xin-Yu Guo.
\newblock Identifying fragility for the stock market: Perspective from the
  portfolio overlaps network.
\newblock {\em Journal of International Financial Markets, Institutions and
  Money}, 62:132--151, 2019.

\bibitem{d2019complex}
Anna~Maria D’Arcangelis, Susanna Levantesi, and Giulia Rotundo.
\newblock A complex networks approach to pension funds.
\newblock {\em Journal of Business Research}, 2019.

\bibitem{borgatti1997network}
Stephen~P Borgatti and Martin~G Everett.
\newblock Network analysis of 2-mode data.
\newblock {\em Social networks}, 19(3):243--270, 1997.

\bibitem{borgatti20092}
Stephen~P Borgatti.
\newblock 2-mode concepts in social network analysis.
\newblock {\em Encyclopedia of complexity and system science}, 6:8279--8291,
  2009.

\bibitem{borgatti2011analyzing}
Stephen~P Borgatti and Daniel~S Halgin.
\newblock Analyzing affiliation networks.
\newblock {\em The Sage handbook of social network analysis}, 1:417--433, 2011.

\bibitem{latapy2008basic}
Matthieu Latapy, Cl{\'e}mence Magnien, and Nathalie Del~Vecchio.
\newblock Basic notions for the analysis of large two-mode networks.
\newblock {\em Social networks}, 30(1):31--48, 2008.

\bibitem{zhou2007bipartite}
Tao Zhou, Jie Ren, Mat{\'u}{\v{s}} Medo, and Yi-Cheng Zhang.
\newblock Bipartite network projection and personal recommendation.
\newblock {\em Physical review E}, 76(4):046115, 2007.

\bibitem{lavin2019modeling}
Jaime~F Lavin, Mauricio~A Valle, and Nicol{\'a}s~S Magner.
\newblock Modeling overlapped mutual funds’ portfolios: A bipartite network
  approach.
\newblock {\em Complexity}, 2019, 2019.

\bibitem{delpini2019systemic}
Danilo Delpini, Stefano Battiston, Guido Caldarelli, and Massimo Riccaboni.
\newblock Systemic risk from investment similarities.
\newblock {\em PloS one}, 14(5), 2019.

\bibitem{10.1145/2939672.2939754}
Aditya Grover and Jure Leskovec.
\newblock Node2vec: Scalable feature learning for networks.
\newblock In {\em Proceedings of the 22nd ACM SIGKDD International Conference
  on Knowledge Discovery and Data Mining}, KDD 2016, page 855–864, New York,
  NY, USA, 2016. Association for Computing Machinery.

\bibitem{NIPS2013_5021}
Tomas Mikolov, Ilya Sutskever, Kai Chen, Greg~S Corrado, and Jeff Dean.
\newblock Distributed representations of words and phrases and their
  compositionality.
\newblock In C.~J.~C. Burges, L.~Bottou, M.~Welling, Z.~Ghahramani, and K.~Q.
  Weinberger, editors, {\em Advances in Neural Information Processing Systems
  26}, pages 3111--3119. Curran Associates, Inc., 2013.

\bibitem{mara2019evalne}
Alexandru Mara, Jefrey Lijffijt, and Tijl De~Bie.
\newblock Evalne: a framework for evaluating network embeddings on link
  prediction.
\newblock {\em arXiv preprint arXiv:1901.09691}, 2019.

\bibitem{macqueen1967}
J.~MacQueen.
\newblock Some methods for classification and analysis of multivariate
  observations.
\newblock In {\em Proceedings of the Fifth Berkeley Symposium on Mathematical
  Statistics and Probability, Volume 1: Statistics}, pages 281--297, Berkeley,
  Calif., 1967. University of California Press.

\bibitem{scikit-learn}
F.~Pedregosa, G.~Varoquaux, A.~Gramfort, V.~Michel, B.~Thirion, O.~Grisel,
  M.~Blondel, P.~Prettenhofer, R.~Weiss, V.~Dubourg, J.~Vanderplas, A.~Passos,
  D.~Cournapeau, M.~Brucher, M.~Perrot, and E.~Duchesnay.
\newblock Scikit-learn: Machine learning in {P}ython.
\newblock {\em Journal of Machine Learning Research}, 12:2825--2830, 2011.

\bibitem{rosenberg2007v}
Andrew Rosenberg and Julia Hirschberg.
\newblock V-measure: A conditional entropy-based external cluster evaluation
  measure.
\newblock In {\em Proceedings of the 2007 joint conference on empirical methods
  in natural language processing and computational natural language learning
  (EMNLP-CoNLL)}, pages 410--420, 2007.

\bibitem{gao2018bine}
Ming Gao, Leihui Chen, Xiangnan He, and Aoying Zhou.
\newblock Bine: Bipartite network embedding.
\newblock In {\em The 41st international ACM SIGIR conference on research \&
  development in information retrieval}, pages 715--724, 2018.

\bibitem{liu2019hood2vec}
Xin Liu, Konstantinos Pelechrinis, and Alexandros Labrinidis.
\newblock hood2vec: Identifying similar urban areas using mobility networks.
\newblock {\em arXiv preprint arXiv:1907.11951}, 2019.

\end{thebibliography}
\bibliographystyle{unsrt}

\end{document}